Extension of Intra-Nuclear Cascade Model
to Neutron Induced Nonelastic Cross-Sections in Low Energy Region

Masahiro Nakano[1,2] and Yusuke Uozumi[2]

[1]Junshin Gakuen University, [2]Kyushu University

[1,2] Fukuoka, Japan

e-mail: nakano@med.uoeh-u.ac.jp

Two features, a slow slope and a sharp drop, in neutron induced total nonelastic cross-sections are analyzed within the framework of an intra-nuclear cascade (INC) model. First, to reproduce the slow slope from 100MeV to 10 MeV, the original INC is generalized in two points; a method to construct the ground state of the target nucleus, and a method of taking the effective two body cross-sections between two nucleons. Secondly, to analyze the origin of the sharp drop from 10MeV to nearly 1 MeV, the INC is extended to include quantum effects which are originated from the existence of the discrete states in the nuclear potential. It is shown that this extension leads to the sharp drops in the very low energy below around 10MeV. It is concluded that the INC model can be extended to explain the sharp drops in addition to the slow slope in neutron induced nonelastic cross-sections in the energy region from 100MeV down to 1 MeV.

## I. INTRODUCTION

Nonelastic cross-section is defined as the total reaction cross-section minus the elastic cross section. It includes all the reactions such as particle emissions and absorptions except the elastic reaction. Experimental data on the neutron induced nonelastic cross-section are very few because of difficulty of the measurement, especially in the low energy region below 100MeV. In fig.1, we show the data of $^{208}$Pb and $^{27}$Al, where nonelastic cross-sections are relatively well measured in a wide range[1-13]. Although the data have experimental errors, the tendency of the cross-section is similar to each other. Two common features are clearly observed in the neutron induced nonelastic cross-section in the energy region less than 100MeV. It shows a gradually rising slope as the incident energy of neutron becomes small in the energy region from 100MeV to around 10MeV, and a sharp drop in a narrow energy range from around 10MeV to nearly 1MeV. It is important that the common features are observed in typical heavy and light nuclei; $^{208}$Pb and $^{27}$Al. Furthermore partial data of several nuclei $_6$C[14], $_{26}$Fe[14,15], $_{83}$Bi[15,16] in the energy region around 10MeV have the same tendency. Therefore it is considered that the remarkable features of a slow slope from 100MeV and a sharp drop from around 10MeV are widely observed.

There is a phenomenological model [17] to reproduce the slow slope from 100MeV to 10MeV. In the model, a transmission probability is calculated by the integration on the path of the injected nucleon through the nucleus, where two nucleon cross-sections and a Wood Saxon density are used. The model could reproduce the data more than 10MeV, however, the reason of sharp drop of the neutron induced cross-section has not been explained.

This sharp drop is one of mysterious problems of the neutron induced nonelastic cross-section. If the projectile is proton, the Coulomb interaction between the proton and target nuclei strongly reject the injection at the very low energy, then it is natural that the proton induced reaction cross-section sharply drops. The Coulomb barrier, however, cannot affect the neutron.

There have been no dynamic models to explain this sharp drops together with the slow slope. Therefore one of important aims of this paper is to analyze the origin of the sharp drop. The other aim is to generalize the INC model to reproduce the slow slop from 100MeV to around 10 MeV. In this paper, we introduce two generalizations to the original INC model to reproduce the slow slope, after two generalizations, we extend the generalized INC model to apply the calculation of cross-sections in the low energy region below around 10MeV and to explain the origin of this sharp drops.

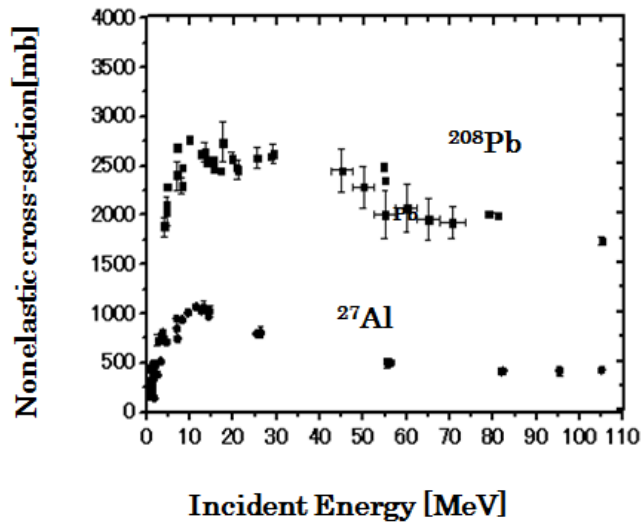

Fig.1 Experimental data of neutron induced nonelastic cross-sections of $^{27}$Al and $^{208}$Pb.

The intra-nuclear cascade (INC) is a model to describe the time development of all the particles in target nuclei and the projectile based on the classical mechanics with relativistic kinematics. In this model, the nonelastic cross-section is counted as sum up the outgoing particles in any energy and any angles. The INC model followed by the generalized evaporation model (GEM) has explained well various reactions such as (p,p'x), (p,dx), (p,αx) in very wide energies and angles [18-21]. On the other hand, the INC model has not been applied to neutron induced reactions in the low energies. A direct application of the original INC cannot reproduce two features, i.e. the slow slope and the sharp drop in the cross-section. Therefore we generalize the original INC in two points. One is a method to construct the ground state of the target nucleus. The generalization of the ground sate brings a better energy dependence of the reaction cross-section. The other is a method of taking the effective two body cross-sections between two nucleons. We explain the two modifications in the following section.

## II. GEBERALIZATION OF THE INC MODEL

### 2.1 Constitution of the ground state

In the original INC, the ground state is prepared based on a random sampling both on the positions and momentums. On the positions of the particles in the nucleus, we use a random number method so as to reproduce the Wood-Saxon density distribution as a whole in a probabilistic way. The densities of the nucleons in the ground states are shown in Fig.2 for $^{27}$AL and $^{208}$Pb. On the other hand, the distribution of momentum in the original INC model is randomly chosen so as to reproduce a uniform distribution which is shown by the dashed line in Fig.3. The total cross-sections calculated using this ground state of the random sampling bring a peak around Ein=40MeV both for Al and Pb which is shown by the broken lines in Fig.4. The tendency is quite different from the experimental data.

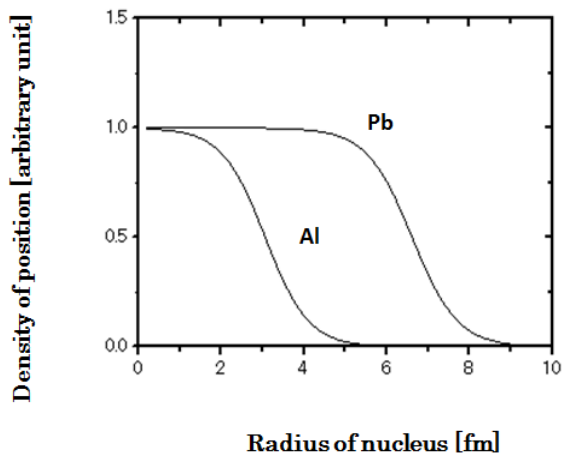

Fig.2 Distributions of positions of nucleons in the ground states for $^{27}$AL and $^{208}$Pb. Wood-Saxon shape is realized as a whole.

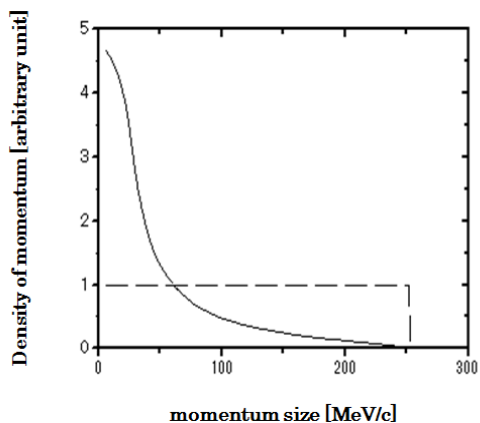

Fig3 Distributions of momentum in the ground states, which are the same for $^{27}$AL and $^{208}$Pb. The original INC model gives the uniform distribution (dashed line), on the other hand, the local dependent momentum method gives a sharply damped distribution (solid line).

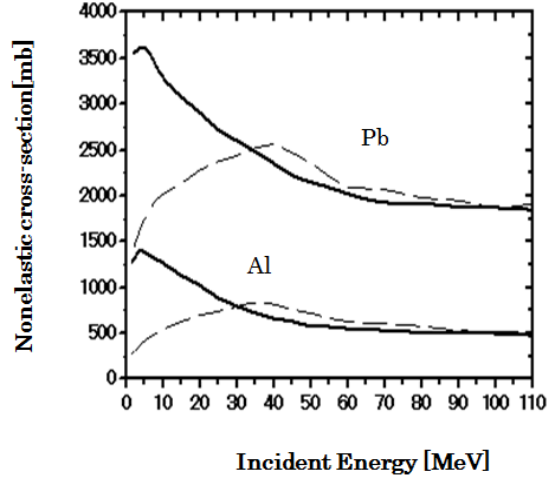

Fig.4 Neutron induced nonelastic cross-sections calculated with the ground state constructed by random method (broken line) and by the new method (solid line). The same two nucleon cross-sections in eqs. (6) –(8) are taken for both calculations.

In this work, we generalize the uniform ground state to a new ground state based on a local dependent momentum method. On the positions, the same procedure is given by the random setup as the original INC model. On the other hand, the momenta are prepared according to the effective nucleon mass at the particle position. The effective nucleon mass is determined by a local dependent formula,

$$M^*(r) = M + U(r) \qquad (1)$$

where M is the nucleon mass and the potential $U(r)$ has a Wood-Saxon shape as follows;

$$U(r) = V_0 / (1 + \exp((r - r_0)/a_0)) \qquad (2)$$

In this paper, the parameters $r_0$ and $a_0$ of the Wood-Saxon are chosen from the experimental data of charge distribution by electron scattering experiments [22], which are given in Table 1.

Table 1. Wood-Saxon parameters, the radius $r_0$ and the diffuseness $a_0$ for two targets $^{27}$Al and $^{208}$Pb.

|  | $r_0$[fm] | $a_0$[fm] |
|---|---|---|
| $^{27}$Al | 2.840 | 0.569 |
| $^{208}$Pb | 6.620 | 0.546 |

Using this effective nucleon mass, the maximum momentum at $r$ is given by

$$P_{\max}(r) = \sqrt{E_f^2 - M^*(r)^2} \qquad (3)$$

where $E_f$ is the Fermi energy given by the nucleon mass minus the binding energy, i.e., $E_f$ = 938.93-8.74 MeV in this paper. The momentum of particles are determined by random numbers, which are chosen in a probabilistic way from zero to the maximum momentum $P_{max}(r)$. As a result, the new ground state has a sharply damped distribution as shown by the solid lone in the Fig.3. It gives smaller momentum in a peripheral region of the nucleus. This new ground state brings a better agreement with the data than the uniform distribution. The calculation gives completely different curvature from the result of the uniform distribution; a slow slope as the incident energy goes to small, which is shown by the solid line in Fig.4. This result indicates that the local dependent distribution should be selected instead of the uniform distribution.

### 2.2 Effective two body cross-sections between two nucleons

Cugnon et al. [23] introduced improved two body cross-sections given by following equations for better fits to low energy phenomena. They insist that these cross-sections were made to reproduce the free NN cross-sections, and is valid down to $p$ =0.1$GeV/c$. The two-body cross-sections give a sufficient fit, especially to higher energy phenomena. The two-body cross-sections $S$[mb] are expressed in the unit of mb by the following equations for each energy interval:

*for pp*

$$\begin{aligned}
S &= 41 + 60(P_G - 0.9)\exp(-1.2 P_G) && for\ 1.5 GeV/c < p_G < 5 GeV/c \\
S &= 23.5 + 24.6/(1 + \exp(-(P_G - 1.2)/0.1)) && for\ 0.8 GeV/c < p_G < 1.5 GeV/c \\
S &= 23.5 + 1000(P_G - 0.7)^4 && for\ 0.4 GeV/c < p_G < 0.8 GeV/c \\
S &= 34(P_G/0.4)^{-2.104} && for\ p_G < 0.4 GeV/c.
\end{aligned}$$
(4)

*for np and nn*

$$\begin{aligned}
S &= 42 && for\ p_G > 2 GeV/c \\
S &= 24.2 + 8.9 P_G && for\ 1 GeV/c < p\_G < 2 GeV/c \\
S &= 33 + 196\,abs(P_G - 0.95)^{2.5} && for\ 0.4 GeV/c < p_G < 1 GeV/c \\
S &= 6.3555 P_G^{-3.2481}\exp(-0.377(\ln P_G)^2) && for\ p_G < 0.4 GeV/c.
\end{aligned}$$
(5)

where $p_G$ is the relative momentum of the two nucleons in the unit of GeV/c.
The two body cross-sections of the improved Cugnon has defects, i.e, there are jumps in the curvature at each edge point since they are separately given for each interval. Furthermore the INC model calculations using the improved two body cross-sections by Cugnon largely overestimate the data as shown in Fig.6. Therefore we introduced a new set of effective two body cross-sections. In the calculations of $^{208}$Pb and $^{27}$Al., we used the same two body cross sections. The expression is not unique since many formula can represent a similar shape.

One formula is given for $p_G < 2 GeV/c$ as follows:

$$S = (Y1+Y2)(0.80+0.20/(1+\exp(-(P_G-0.8)/0.3)))+0.31 Y3 \quad (6)$$

where the functions $Y1, Y2$ and $Y3$ are functions given as follows:

for pp

$$Y1 = 250\exp(-P_G^{1.2}/0.1)$$
$$Y2 = 26.5/(1+\exp(-(P_G-1.178)/0.122))+22$$
$$Y3 = 3300\exp(-P_G^{0.8}/0.07)+80000\exp(-P_G^{0.86}/0.02)$$

$$(7)$$

for np and nn

$$Y1 = 67\exp(-(P_G-0.12)^2/0.15))$$
$$Y2 = (10P_G+23)(1+0.2\exp(-(P_G-0.5)/0.15))$$
$$Y3 = 8000\exp(-P_G/0.064)$$

$$(8)$$

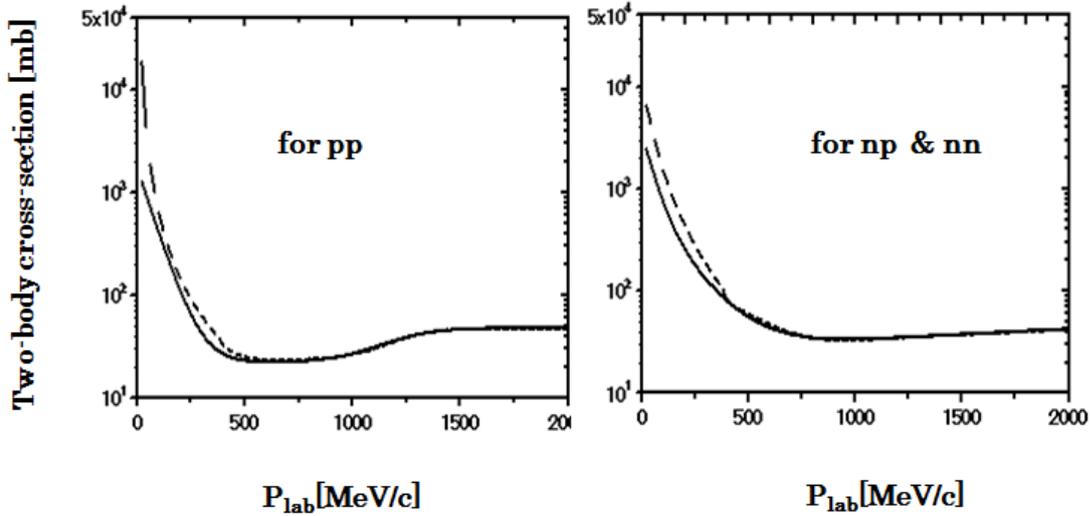

Fig.5 Two body reaction cross-section for pp(left) and pn & nn(right) by the improved Cugnon (dashed line) in eqs.(4),(5) and the proposed one (solid line) in eqs.(6)-(8).

As illustrated in Fig.5, the proposed two body cross-sections are similar to those to the improved Cugnon et al.. Our two body cross-sections are slightly smaller than those by the improved Cugnon in the momentum range smaller than $p_G = 1.0$GeV/c. This implies that the free cross-sections should be reduced as a result of the medium effects in the nuclear matter.

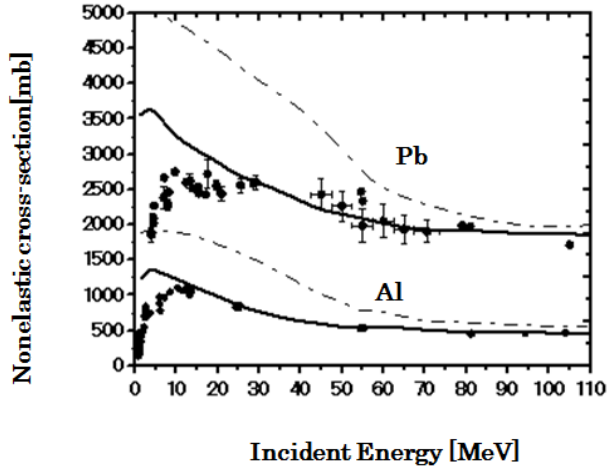

Fig.6. Comparison between the Experimental data(dots) and two results by using the improved two body cross-sections (dash-dotted lines) in eqs.(4),(5) and by the proposed cross-sections (solid lines) in eqs.(6)-(8) for $^{27}$Al and $^{208}$Pb.

The calculated result based on the two generalizations reproduces the slow slope in the experimental data as is shown in Fig.6. On the other hand, it cannot reproduce the sharp drop below around 10MeV. The slow slope of the calculated cross-section results from the fact that the two body cross-sections are sharply rising in the low energy as shown in Fig.5. Hence it is obvious that a further extension is necessary for the generalized INC to describe the sharp drop of the experimental data in the low energy less than around 10MeV.

### III. EXTENSION of THE INC MODEL

The INC model is based on classical dynamics. Then there is a limitation of a classical model, that is, a simple model cannot include quantum effects. There is an important quantum effect, i.e., effects originated from Pauli principle. In the original INC model, the effects have been already included effectively by the treatment that the process is forbidden if the energy of a scattered particle is below the Fermi sea. The importance of the Pauli principle is also pointed out in the phenomenological model mentioned above [6].

In very low energies, there is another important quantum effect, that is originated from the existence of discrete states. Until now, the INC model treats the states of the scattered particles in the nuclear potential as continuous states. In quantum mechanics, scattered particles go to the bound states, whose energies are quantized. We simulate this effect as follows. The transition in energy of scattered two particles is

$$E1 + E2 \rightarrow E1' + E2' \qquad (9)$$

The momentum and energy is conserved in the INC model after a collision, so that $E1+E2 = E1'+E2'$.

The probability of the two nuclei is the multiplication of the probabilities of two nucleons;

$$P(E1', E2') = P(E1')P(E2') \tag{10}$$

The transition probability of one particle is

$$P(E) = \sum_i \exp(-((E-E_i)/a_i)^2) \tag{11}$$

where the index $i$ in the summation is the number of bound states.

The transition probability $P(E)=1$ when the energy is over the free energy, that is the $E_{Fermi}$ + Binding energy for neutron, and $E_{Fermi}$ + Binding energy + Coulomb barrier for proton. The binding energy is set to be 8.74MeV for both nuclei, and Coulomb barrier is 3.30MeV for $^{27}$Al, and 8.92 MeV for $^{208}$Pb respectively.

This condition on the free energy is important for the nucleon over the critical energy to propagate without no restriction, that corresponds to just classical situation.

The levels are listed for neutron and proton of $^{27}$Al in the table2 and $^{208}$Pb in the table3 [24]. Owing to the Coulomb barrier, more proton levels are included than neutron levels. The energies of single particle level $E_i$ are slightly adjusted to reproduce the data. It is noteworthy that the sharp drop in the experimental data does not depend on the detail of the parameters.

Table 2. Energies of single particle levels for neutron and proton of Al. The listed energies are measured from the Fermi sea level. The symbols start from n(neutron) or p(proton), and show the principal quantum number, angular momentum, and total spin of the state [13].

| n2s1/2 | n1d3/2 | n1f7/2 | n2p3/2 | p2s1/2 | p1d3/2 | p1f7/2 | p2p3/2 | p1f5/2 | p2p1/2 |
|---|---|---|---|---|---|---|---|---|---|
| 3.5 | 5 | 7 | 8 | 3.5 | 5 | 7 | 8 | 10 | 12 |

Table3. Energies of single particle levels for neutron and proton of Pb. Captions are the same as Table 2.

| n2g9/2 | n1i11/2 | n3d5/2 | p2f7/2 | p1h9/2 | p1i13/2 | p3p3/2 |
|---|---|---|---|---|---|---|
| 6 | 10 | 15 | 7 | 8 | 12 | 17 |

The constant $a_i$ is the width of the probability, and is set to proportional to the spin multiplicity $(2j+1)$.

$$a_i = (2j+1)\, h \tag{12}$$

where the width $h$ is set to be 0.18 MeV.

We require that two nucleons having the energies E1' and E2' follow the probability distribution as a whole. Actually we realize this transition probability in a stochastic way. After the energy of two particles E1', E2' are determined in a virtual collision, we compare a random number $r_i$ and the probability $P(Ei')$, then the process is forbidden when the random number $r_i$ is larger than the transition probability $P(Ei')$.

## IV. RESULTS AND DISCUSSIONS

The INC model as extended in this way descries excellently the sharp drops both in $^{208}$Pb and $^{27}$Al as shown in Fig.7. The reason of the discrepancy between the generalized INC by the solid lines and the experimental data in Fig.6 is now clear. The origin of the sharp drop is from the quantum effect. The fact that the scattered two particles should go to the discrete bound states confines the phase space in energy of the two nucleons. The phase space goes to narrow as the energy of the induced neutron goes to small. This is a kind of bound state constriction.

In the case of the large injection energy, the scattered nucleon inside the nucleus goes up to a sufficiently high energy than the free energy, then $P(Ei')=1$ for continues energies. In this case, there are no quantum effects, then a classical model works effectively. On the other hand, in the case of a very low injection energy, the particle moves to a little over Fermi sea, and the energy of the injected neutron goes down near the Fermi sea, where the transition probability P(Ei') is nearly zero since there is no allowed state. This means that most of these transitions cannot be allowed based on the quantum mechanics. This leads to the sharp drop in the cross sections in very low energy. It is natural that this drop occurs especially in the very low energy region.

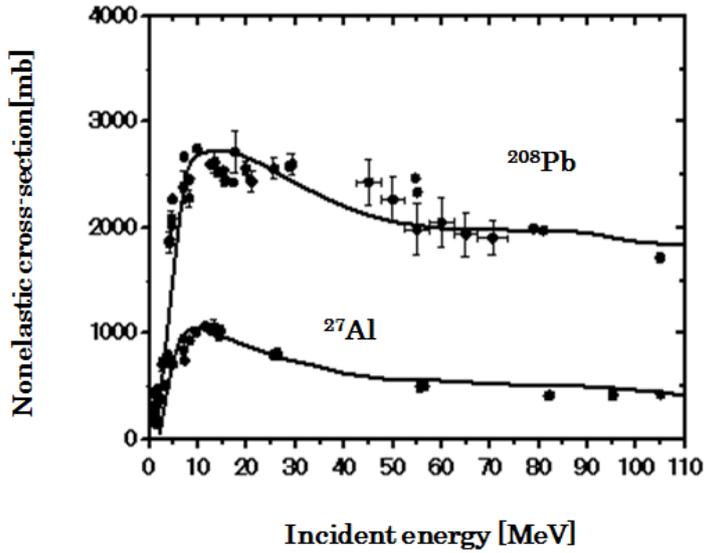

Fig.7. Comparison between the Experimental data(dots) and the calculated result by the extended INC (solid lines) for $^{27}$Al and $^{208}$Pb.

We introduced the transition probability $P(E1)$ to represent a quantum effect. Using the transition probability $P(E1)$, we can include the effect of Pauli principle easily. It is sufficient that we set the condition on the transition probability $P(E1) = 0$ for $E1 < E_{Fermi}$. In this way, the two quantum effects are described in a uniformed way.

V. CONCLUSIONS

We have generalized the original INC model in two points; the ground state and the two-nucleon cross-sections, and these generalizations bring a fit to the slow slope in the neutron induced cross-sections.

In addition to these two generalizations, we further extend the generalized INC to include the effect originated from quantum mechanics.

There are two quantum effects in the INC model. The one is the effects from Pauli principle, which has been already included in the original INC model. The other quantum effect is newly included one which is originated from "bound state constraint". The origin of the sharp drop is that the allowed phase space of scattered two nucleons becomes narrow as the energy of the induced neutron goes to very low. It is noted that this effect is confined in the very low energy, then the original INC works well in a wide range of high energy region.

We have shown that the sharp drops are well reproduced in $^{27}$Al and $^{208}$Pb targets by the extended INC model. It is noted that the experimental data in $^{27}$Al and $^{208}$Pb are reproduced by using the same two nucleon cross-sections which are modified slightly by the medium effect from the free ones. The effect is considered not to depend on the target mass. It is noteworthy that the sharp drops in the calculations does not depend on the detail of the parameters of the single particle levels.

Through our analysis, conclusions are (1) the slow slope from 100MeV to around 10MeV is reproduced by two generalizations, (2) the sharp drops of the nonelastic reaction cross section in very low energy region from around 10MeV is explained by quantum effects due to the existence of discrete states, (3) finally the INC model can be extended to include the quantum effects, and owing to this extension, the INC model can well describe the experimental data from 100MeV down to around 1 MeV.

ACKNOWLEDGMENTS

We are grateful for Dr. Yuji Yamaguchi and Dr. Gaku Watanabe, who continuously supported this work.

______________________________________________